\documentclass[a4paper]{jpconf} 
\usepackage{graphicx}
\usepackage{icsghead}

\usepackage{amsmath}
\usepackage{amssymb}
\usepackage{cite}

\begin{document}
\title{Single-trial estimation of stimulus and spike-history effects on time-varying ensemble spiking activity of multiple neurons: a simulation study}
\author{Hideaki Shimazaki}
\address{RIKEN Brain Science Institute, Wako, Saitama, Japan}
\ead{shimazaki@brain.riken.jp}

\begin{abstract}
Neurons in cortical circuits exhibit coordinated spiking activity,
and can produce correlated synchronous spikes during behavior and
cognition. We recently developed a method for estimating the dynamics
of correlated ensemble activity by combining a model of simultaneous
neuronal interactions (e.g., a spin-glass model) with a state-space
method (Shimazaki et al. 2012 PLoS Comput Biol 8 e1002385). This method
allows us to estimate stimulus-evoked dynamics of neuronal interactions
which is reproducible in repeated trials under identical experimental
conditions. However, the method may not be suitable for detecting
stimulus responses if the neuronal dynamics exhibits significant variability
across trials. In addition, the previous model does not include effects
of past spiking activity of the neurons on the current state of ensemble
activity. In this study, we develop a parametric method for simultaneously
estimating the stimulus and spike-history effects on the ensemble
activity from single-trial data even if the neurons exhibit dynamics
that is largely unrelated to these effects. For this goal, we model
ensemble neuronal activity as a latent process and include the stimulus
and spike-history effects as exogenous inputs to the latent process.
We develop an expectation-maximization algorithm that simultaneously
achieves estimation of the latent process, stimulus responses, and
spike-history effects. The proposed method is useful to analyze an
interaction of internal cortical states and sensory evoked activity. 
\end{abstract}

\section{Introduction}

Neurons in the brain make synaptic contacts to each other and form
specific signaling networks. A typical cortical neuron receives synaptic
inputs from $3000-10000$ other neurons, and makes synaptic contacts
to several thousands of other neurons. They send and receive signals
using pulsed electrical discharges known as action potentials, or
spikes. Therefore, individual neurons in a circuit can be activated
in a coordinated manner when relevant information is processed. In
particular, nearly simultaneous spiking activity of multiple neurons
(synchronous spikes) occurs dynamically in relation to a stimulus
presented to an animal, the animal's behavior, and the internal state
of the brain (attention and expectation) \cite{Vaadia95,Riehle97,Steinmetz00,Fujisawa08,Ito10}. 

Recently, it was reported that a model of synchronous spiking activity
that accounts for spike rates of individual neurons and interactions
between pairs of neurons can explain $\sim90$\% of the synchronous
spiking activity of a small subset ($\sim10$) of retinal ganglion
cells \cite{Schneidman06,Shlens06} and cortical neurons \cite{Tang08}
\textit{in vitro}. This model is known as a maximum entropy model
or an Ising/spin-glass model in statistical physics. However, since
the model assumes stationary data, it is not directly applicable to
non-stationary data recorded from awake behaving animals. In these
data sets, spike-rates of individual neurons and even interactions
among them may vary across time. 

In order to analyze time-dependent synchronous activity of neurons,
we recently developed a method for estimating the dynamics of correlations
between neurons by combining the model of neuronal interactions (e.g.,
the Ising/spin-glass model) with a state-space method \cite{Shimazaki09,Shimazaki12}.
In classical neurophysiological experiments, neuronal activity is
repeatedly recorded under identical experimental conditions in order
to obtain reproducible features in the spiking activity across the
`trials'. Typically, neurophysiologists estimate average time-varying
firing rates of individual neurons in response to a stimulus from
the repeated trials \cite{Shimazaki07a,Shimazaki10}. In the same
fashion, the state-space method in \cite{Shimazaki09,Shimazaki12}
aims to estimate the dynamics of the neuronal interactions, including
higher-order interactions, that occurs repeatedly upon the onset of
externally triggered events. When this method is applied to three
neurons recorded simultaneously from the primary motor cortex of a
monkey engaged in a delayed motor task (data from \cite{Riehle97}),
it was revealed that these neurons dynamically organized into a group
characterized by the presence of a higher-order (triple-wise) interaction,
depending on the behavioral demands to the monkey \cite{Shimazaki10}. 

However, neurophysiological studies in the past decades revealed that
spiking activity of individual neurons is subject to large variability
across trials due to structured ongoing activity of the networks that
arises internally to the brain \cite{Arieli96,Tsodyks99,Kenet03}.
In these conditions, the method developed in \cite{Shimazaki10} would
not efficiently detect the stimulus responses because a signal-to-noise
ratio may be small even in the trial-averaged activity. Although statistical
methods for detecting responses of individual neurons from single-trial
data have been investigated \cite{Nawrot99,Cunningham08,Czanner08,Yu09},
no methods are available for estimating synchronous responses of multiple
neurons to a stimulus in a single trial when these neurons are subject
to the activity that is largely unrelated to the stimulus.

In the analysis of single-trial data, it is critical to consider dependency
of the current activity of neurons on the past history of their activity.
A neuron undergoes an inactivation period known as a refractory period
after it generates an action potential. Therefore, a model neuron
significantly improves its goodness-of-fit to data if it captures
this biophysical property \cite{Brown01,Barbieri01}. In addition,
estimating the dependency of the current activity level of a neuron
on past spiking history of another neuron allows us to construct effective
connectivity of the network within an observed set of neurons \cite{Truccolo05,Pillow08}.
Including the spike-history effects in the models of synchronous ensemble
activity is thus an important topic, and investigated also in \cite{Kass11}
in the framework of a continuous-time point process theory. 

In this study we construct a method for simultaneously estimating
the stimulus and spike-history effects on ensemble spiking activity
when the activity of these neurons is dominated by ongoing activity.
For this goal, we extend the previously developed state-space model
of neuronal interactions: We model the ongoing activity, i.e., time-varying
spike rates and interactions, of neurons as a latent process, and
include the stimulus and spike-history effects on the activity as
exogenous inputs to the latent process. We develop an expectation-maximization
(EM) algorithm for this model, which efficiently combines construction
of a posterior density of the latent process and estimation of the
parameters for stimulus and spike-history effects. The method is tested
using simulated spiking activity of 3 neurons with known underlying
architecture. We provide an approximation method for determining inclusion
of these exogenous inputs in the model and a surrogate method to test
significance of the estimated parameters.

\section{Methods}

In this study, we analyze spike sequences simultaneously obtained
from $N$ neurons. From these spike sequences, we construct binary
spike patterns at discrete time steps by dividing the sequences into
disjoint time bins with an equal width of $\Delta$ ms (in total,
$T$ bins). The width $\Delta$ determines a permissible range of
synchronous activity of neurons in this analysis. We let $X_{i}^{t}$
be a binary variable of the $i$-th neuron ($i=1,2,\ldots,N$) in
the $t$-th time bin ($t=1,2,\ldots,T$). Here a time bin containing
`$1$' indicates that one or more spikes exist in the time bin whereas
`$0$' indicates that no spike exists in the time bin. The binary
pattern of $N$ neurons at time bin $t$ is denoted as $\mathbf{X}_{t}=[X_{1}^{t},X_{2}^{t},\ldots,X_{N}^{t}]'$.
The prime indicates the transposition operation to the vector. The
entire observation of the discretized ensemble spiking activity is
represented as $\mathbf{X}_{1:T}=[\mathbf{X}_{1},\mathbf{X}_{2},\ldots,\mathbf{X}_{T}]$.

\subsection{The model of time-varying simultaneous interactions of neurons}

We analyze the ensemble spike patterns using time-dependent formulation
of a joint probability mass function for binary random variables.
Let $x_{i}$ be a binary variable, namely $x_{i}=\left\{ 0,1\right\} $.
The joint probability mass function of $N$-tuple binary variables,
$\mathbf{x}=[x_{1},x_{2},\ldots,x_{N}]$, at time bin $t$ ($t=1,2,\ldots,T$)
can be written in an exponential form as 
\begin{align}
p(\mathbf{x}|\boldsymbol{\theta}_{t}) & =\exp\left[\sum_{i}\theta_{i}^{t}x_{i}+\sum_{i<j}\theta_{ij}^{t}x_{i}x_{j}+\cdots+\theta_{1\cdots N}^{t}x_{1}\cdots x_{N}-\psi(\boldsymbol{\theta}_{t})\right].\label{eq:log-linear}
\end{align}
Here $\boldsymbol{\theta}_{t}=[\theta_{1}^{t},\theta_{2}^{t},\ldots,\theta_{12}^{t},\theta_{13}^{t},\ldots,\theta_{1\cdots N}^{t}]'$
summarizes the time-dependent canonical parameters of the exponential
family distribution. The canonical parameters for the interaction
terms, e.g., $\theta_{ij}^{t}$ ($i,j=1,\ldots,N$), represent time-dependent
instantaneous interactions at time bin $t$ among the neurons denoted
in its subscript. $\psi(\boldsymbol{\theta}_{t})$ is a log normalization
parameter to satisfy $\sum p(\mathbf{x}|\boldsymbol{\theta}_{t})=1$. 

Using a feature vector that captures simultaneous spiking activities
of subsets of the neurons, $\mathbf{f}=[f_{1},f_{2},\ldots,f_{12},f_{13},\ldots,f_{1\cdots N}]'$,
where

\[
\begin{array}{cc}
f_{i}\left(\mathbf{x}\right)=x_{i}, & i=1,\cdots,N\\
f_{ij}\left(\mathbf{x}\right)=x_{i}x_{j}, & i<j\\
\vdots & \mbox{}\\
f_{1\cdots N}\left(\mathbf{x}\right)=x_{1}\cdots x_{N}, & \mbox{}
\end{array}
\]
the probability mass function (Eq.~\ref{eq:log-linear}) is compactly
written as $p({\mathbf{x}}|\boldsymbol{\theta}_{t})=\exp\left[\boldsymbol{\theta}_{t}'\mathbf{f}\left(\mathbf{x}\right)-\psi(\boldsymbol{\theta}_{t})\right]$.
The expected occurrence rates of simultaneous spikes of multiple neurons
is given by a vector $\boldsymbol{\eta}_{t}=E\left[\mathbf{f}\left(\mathbf{x}\right)|\boldsymbol{\theta}_{t}\right]$,
where expectation is performed using $p({\mathbf{x}}|\boldsymbol{\theta}_{t})$.

Eq.~\ref{eq:log-linear} specifies the probabilities of all $2^{N}$
spike patterns by using $2^{N}-1$ parameters. One reasonable approach
to reduce the number of parameters is to select and fix interesting
features in the spiking activity, and construct a probability model
that maximizes entropy. For example, maximization of entropy of the
spike patterns given the low-order features, $\mathbf{f}=[f_{1},f_{2},\ldots,f_{N},f_{12},f_{13},\ldots,f_{N-1,N}]'$,
yields a spin-glass model that is similar to Eq.~\ref{eq:log-linear},
but does not include interactions higher than the second order. While
it is important to explore a characteristic feature vector to neuronal
ensembles, here we note that the method developed in this study does
not depend on the choice of the vector, $\mathbf{f}$. Below, we denote
$d$ as the number of elements in the vector, $\mathbf{f}$.

Given the observed ensemble spiking activity $\mathbf{X}_{1:T}$,
the likelihood function of $\boldsymbol{\theta}_{1:T}=[\boldsymbol{\theta}_{1},\boldsymbol{\theta}_{2},\ldots,\boldsymbol{\theta}_{T}]$
is given as 
\begin{align}
p\left(\mathbf{X}_{1:T}|\boldsymbol{\theta}_{1:T}\right) & =\prod\limits _{t=1}^{T}\exp[\boldsymbol{\theta}_{t}'\mathbf{f}\left(\mathbf{X}_{t}\right)-\psi(\boldsymbol{\theta}_{t})],\label{eq:observation_equation}
\end{align}
assuming conditional independence across the time bins. Eq.\ref{eq:observation_equation}
constitutes an observation equation of our state-space model.

\subsection{Inclusion of stimulus and spike-history effects in the state model}

The main focus of attention in this study is modeling of a process
for the time-dependent canonical parameters, $\boldsymbol{\theta}_{t}$,
in Eq.~\ref{eq:log-linear}. We model their evolution as a first-order
auto-regressive (AR) model. The effects of the stimulus and spike
history are included as exogenous inputs to the AR model (an ARX model).
In its full expression, the state model is written as 
\begin{equation}
\boldsymbol{\theta}_{t}=\mathbf{F}\boldsymbol{\theta}_{t-1}+\mathbf{G}\mathbf{S}_{t}+\sum_{i=1}^{p}\mathbf{H}_{i}\mathbf{X}_{t-i}+\boldsymbol{\xi}_{t},\label{eq:state_equation}
\end{equation}
for $t=2,\ldots,T$. Here the matrix $\mathbf{F}$ ($d\times d$ matrix)
is the first order auto-regressive parameter. $\boldsymbol{\xi}_{t}$
($d\times1$ matrix) is a random vector independently drawn from a
zero-mean multivariate normal distribution with covariance matrix
$\mathbf{Q}$ ($d\times d$ matrix) at each time bin. The state process
starts with an initial value $\boldsymbol{\theta}_{1}$ that follows
a normal distribution with mean $\boldsymbol{\mu}$ ($d\times1$ matrix)
and covariance matrix $\boldsymbol{\Sigma}$ ($d\times d$ matrix),
namely $\boldsymbol{\theta}_{1}\sim\mathcal{N}\left(\boldsymbol{\mu},\boldsymbol{\Sigma}\right)$.
Below, we describe details of the exogenous terms. 

The second term represents responses to external signals, or stimuli,
$\mathbf{S}_{t}$, which are observed concurrently with the spike
sequences. The vector $\mathbf{S}_{t}$ is a column vector of $n_{s}$
external signals at time bin $t$. The each element is the value of
an external signal at time bin $t$. If an external signal is represented
as a sequence of discrete events, we denote the corresponding element
of $\mathbf{S}_{t}$ by `1' if an event occurred within time bin $t$
and `0' otherwise. Multiplying $\mathbf{S}_{t}$ by the matrix $\mathbf{G}$
($d\times n_{s}$ matrix) produces weighted linear combinations of
the external signals at time bin $t$.

The third term represents the effects of spiking activity during the
previous $p$ time bins, $\mathbf{X}_{t-i}$ $(i=1,\ldots,p)$, on
the current activity. The matrix $\mathbf{H}_{i}$ ($d\times N$ matrix)
represents the spike-history effects of spiking activity in the previous
time bin $t-i$ on the state at time bin $t$. The spike-history effects
are collectively denoted as $\mathbf{H}\equiv[\mathbf{H}_{1},\mathbf{H}_{2},\ldots,\mathbf{H}_{p}]$
($d\times Np$ matrix). 

Eq.~\ref{eq:state_equation} constitutes a prior density of the latent
process in our state-space model. We denote the set of parameters
in the prior distribution, called hyper-parameters, as $\mathbf{w}\equiv\left[\mathbf{F},\mathbf{G},\mathbf{H},\mathbf{Q},\boldsymbol{\mu},\boldsymbol{\Sigma}\right]$.
In this study, we refer to $\mathbf{w}$ as a parameter. In addition,
we simplify Eq.~\ref{eq:state_equation} as $\boldsymbol{\theta}_{t}=\mathbf{F}\boldsymbol{\theta}_{t-1}+\mathbf{U}\mathbf{u}_{t}+\boldsymbol{\xi}_{t},$
where $\mathbf{u}_{t}$ is a single column vector constructed by stacking
the stimulus vector and spike-history vectors at time bin $t$ in
a row, i.e., $\mathbf{u}_{t}=[\mathbf{S}_{t};\mathbf{X}_{t-1};\mathbf{X}_{t-2};\ldots;\mathbf{X}_{t-p}]$.
Similarly, we define a matrix $\mathbf{U}$ as $\mathbf{U}=[\mathbf{G},\mathbf{H}]$.
With this simplification, the prior density defined in Eq.~\ref{eq:state_equation}
is written as $p(\boldsymbol{\theta}_{1:T}|\mathbf{w})=p(\boldsymbol{\theta}_{1}|\boldsymbol{\mu},\boldsymbol{\Sigma})\prod_{t=2}^{T}p(\boldsymbol{\theta}_{t}|\boldsymbol{\theta}_{t-1},\mathbf{F},\mathbf{U},\mathbf{Q})$,
where the transition probability, $p(\boldsymbol{\theta}_{t}|\boldsymbol{\theta}_{t-1},\mathbf{F},\mathbf{U},\mathbf{Q})$,
is given as a normal distribution with mean $\mathbf{F}\boldsymbol{\theta}_{t-1}+\mathbf{U}\mathbf{u}_{t}$
and covariance matrix $\mathbf{Q}$.

\section{Estimation of stimulus responses and spike-history effects}

We estimate the parameter, $\mathbf{w}$, based on the principle of
maximizing a (log) marginal likelihood function. Namely, we select
the parameter that maximizes 
\begin{equation}
l\left(\mathbf{w}\right)=\log\int p\left(\mathbf{X}_{1:T},\boldsymbol{\theta}_{1:T}|\mathbf{w}\right)d\boldsymbol{\theta}_{1:T}.\label{eq:marginal_loglikelihood}
\end{equation}
For this goal, we use the expectation-maximization (EM) algorithm
\cite{Dempster77,Shumway82,Smith03}. In this method, we iteratively
obtain the optimal parameter $\mathbf{\mathbf{w}^{\ast}}$ that maximizes
the lower bound of the above log marginal likelihood. This alternative
function, known as the expected complete data log-likelihood (a.k.a.,
$q$-function), is computed as 
\begin{align}
q\left(\mathbf{\mathbf{w}^{\ast}}|\mathbf{w}\right) & \equiv E\left[\log p\left(\mathbf{X}_{1:T},\boldsymbol{\mathbf{\theta}}_{1:T}|\mathbf{\mathbf{w}^{\ast}}\right)|\mathbf{X}_{1:T},\mathbf{w}\right]\nonumber \\
 & =\sum\limits _{t=1}^{T}\left(E\boldsymbol{\theta}'_{t}\mathbf{f}\left(\mathbf{X}_{t}\right)-E\psi\left(\boldsymbol{\theta}_{t}\right)\right)-\frac{d}{2}\log{2\pi}-\frac{1}{2}\log{\det\boldsymbol{\Sigma}^{\ast}}\nonumber \\
 & -\frac{1}{2}E[\left(\boldsymbol{\theta}_{1}-\boldsymbol{\mu}^{\ast}\right)^{\prime}\boldsymbol{\Sigma}^{\ast-1}\left(\boldsymbol{\theta}_{1}-\boldsymbol{\mu}^{\ast}\right)]-\frac{\left(T-1\right)d}{2}\log{2\pi}-\frac{\left(T-1\right)}{2}\log{\det\mathbf{Q}^{\ast}}\nonumber \\
 & -\frac{1}{2}\sum\limits _{t=2}^{T}E[\left(\boldsymbol{\theta}_{t}-\mathbf{F}^{\ast}\boldsymbol{\theta}_{t-1}-\mathbf{U}^{\ast}\mathbf{u}_{t}\right)^{\prime}\mathbf{Q}^{\ast-1}\left(\boldsymbol{\theta}_{t}-\mathbf{F}^{\ast}\boldsymbol{\theta}_{t-1}-\mathbf{U}^{\ast}\mathbf{u}_{t}\right)].\label{eq:Q-function}
\end{align}
The expectation, $E[\centerdot|\mathbf{X}_{1:T},\mathbf{w}]$, in
Eq.~\ref{eq:Q-function} is performed using the smoother posterior
density of the state obtained by a nominal parameter, $\mathbf{w}$,
namely 

\begin{equation}
p\left(\boldsymbol{\theta}_{1:T}|\mathbf{X}_{1:T},\mathbf{w}\right)=\frac{p\left(\mathbf{X}_{1:T}|\boldsymbol{\theta}_{1:T}\right)p\left(\boldsymbol{\theta}_{1:T}|\mathbf{w}\right)}{p\left(\mathbf{X}_{1:T}|\mathbf{w}\right)}.\label{eq:posterior}
\end{equation}
In particular, Eq.~\ref{eq:Q-function} can be computed using the
following expected values by the posterior density: The smoother mean
$\boldsymbol{\theta}_{t|T}=E\left[\boldsymbol{\theta}_{t}|\mathbf{X}_{1:T},\mathbf{w}\right]$,
the smoother covariance matrix $W_{t|T}=E[(\boldsymbol{\theta}_{t}-\boldsymbol{\theta}_{t|T})(\boldsymbol{\theta}_{t}-\boldsymbol{\theta}_{t|T})^{\prime}|\mathbf{X}_{1:T},\mathbf{w}]$,
and the lag-one covariance matrix, $W_{t,t-1|T}=E[(\boldsymbol{\theta}_{t}-\boldsymbol{\theta}_{t|T})(\boldsymbol{\theta}_{t-1}-\boldsymbol{\theta}_{t-1|T})^{\prime}|\mathbf{X}_{1:T},\mathbf{w}]$.
These values are obtained using the approximate recursive Bayesian
filtering/smoothing algorithm developed in \cite{Shimazaki09,Shimazaki12}
(See Appendix A and Eqs.~\ref{eq:smooth_mean}, \ref{eq:smooth_covariance},
and \ref{eq:lag-one_covariance} therein).

In the EM-algorithm, we obtain the parameter that maximizes the $q$-function
by alternating the expectation (E) and maximization (M) steps. In
the E-step, we obtain the above expected values in Eq.~\ref{eq:Q-function}
by the approximate recursive Bayesian filtering/smoothing algorithm
using a fixed $\mathbf{w}$ (Appendix A). In the M-step, we obtain
the parameter, $\mathbf{\mathbf{w}^{\ast}}$, that maximizes Eq.~\ref{eq:Q-function}.
The resulting $\mathbf{\mathbf{w}^{\ast}}$ is then used in the next
E-step. Below, we derive an algorithm for optimizing the parameter
at the M-step. 

For the state model that includes the auto-regressive parameter and
stimulus and/or spike-history effects, these parameters are estimated
simultaneously. From $\frac{\partial}{\partial\mathbf{F}^{\ast}}q\left(\mathbf{w}^{\ast}|\mathbf{w}\right)=\mathbf{0}$,
we obtain 

\begin{equation}
\mathbf{F}^{\ast}\sum_{t=2}^{T}\left(\mathbf{W}_{t-1,t|T}+\boldsymbol{\theta}_{t-1|T}\boldsymbol{\theta}_{t-1|T}^{\prime}\right)+\mathbf{U}^{\ast}\sum_{t=2}^{T}\boldsymbol{u}_{t}\boldsymbol{\theta}_{t-1|T}^{\prime}=\sum_{t=2}^{T}\left(\mathbf{W}_{t-1,t|T}+\boldsymbol{\theta}_{t|T}\boldsymbol{\theta}_{t-1|T}^{\prime}\right).
\end{equation}
Here, $\boldsymbol{\theta}_{t|T}$, $\mathbf{W}_{t|T}$, and $\mathbf{W}_{t-1,t|T}$
are the smoother mean and covariance, and the lag-one covariance matrix
given by Eqs.~\ref{eq:smooth_mean}, \ref{eq:smooth_covariance},
and \ref{eq:lag-one_covariance}, respectively. Similarly, from $\frac{\partial}{\partial\mathbf{U}^{\ast}}q\left(\mathbf{w}^{\ast}|\mathbf{w}\right)=\mathbf{0}$,
we obtain 
\begin{equation}
\mathbf{F}^{\ast}\sum_{t=2}^{T}\boldsymbol{\theta}_{t-1|T}\boldsymbol{u}_{t}^{\prime}+\mathbf{U}^{\ast}\sum_{t=2}^{T}\boldsymbol{u}_{t}\boldsymbol{u}_{t}^{\prime}=\sum_{t=2}^{T}\boldsymbol{\theta}_{t|T}\boldsymbol{u}_{t}^{\prime}.
\end{equation}
Hence, the simultaneous update rule for $\mathbf{F}^{\ast}$ and $ $$\mathbf{U}^{\ast}$
is given as 
\begin{align}
\left[\begin{array}{cc}
\mathbf{F}^{\ast} & \mathbf{U}^{\ast}\end{array}\right] & =\left[\begin{array}{cc}
\sum_{t=2}^{T}\left(\mathbf{W}_{t-1,t|T}+\boldsymbol{\theta}_{t|T}\boldsymbol{\theta}_{t-1|T}^{\prime}\right) & \sum_{t=2}^{T}\boldsymbol{\theta}_{t|T}\boldsymbol{u}_{t}^{\prime}\end{array}\right]\nonumber \\
 & \left[\begin{array}{cc}
\sum_{t=2}^{T}\left(\mathbf{W}_{t-1,t|T}+\boldsymbol{\theta}_{t-1|T}\boldsymbol{\theta}_{t-1|T}^{\prime}\right) & \sum_{t=2}^{T}\boldsymbol{\theta}_{t-1|T}\boldsymbol{u}_{t}^{\prime}\\
\sum_{t=2}^{T}\boldsymbol{u}_{t}\boldsymbol{\theta}_{t-1|T}^{\prime} & \sum_{t=2}^{T}\boldsymbol{u}_{t}\boldsymbol{u}_{t}^{\prime}
\end{array}\right]^{-1}.\label{eq:update_F_U}
\end{align}
Here the inverse matrix on the r.h.s. is obtained by using the blockwise
inversion formula:
\[
\left[\begin{array}{cc}
A & B\\
C & D
\end{array}\right]^{-1}=\left[\begin{array}{cc}
A^{-1}+A^{-1}B\left(D-CA^{-1}B\right)^{-1}CA^{-1} & -A^{-1}B\left(D-CA^{-1}B\right)^{-1}\\
-\left(D-CA^{-1}B\right)^{-1}CA^{-1} & \left(D-CA^{-1}B\right)^{-1}
\end{array}\right].
\]

The covariance matrix, $\mathbf{Q}$, can be optimized separately.
From $\frac{\partial}{\partial\mathbf{Q}^{\ast}}q\left(\mathbf{w}^{\ast}|\mathbf{w}\right)=\mathbf{0}$,
the update rule of $\mathbf{Q}$ is obtained as 
\begin{align}
\mathbf{Q}^{\ast}= & \frac{1}{T-1}\sum_{t=2}^{T}[\mathbf{W}_{t|T}-\mathbf{W}_{t-1,t|T}\mathbf{F}^{\prime}-\mathbf{FW}_{t-1,t|T}^{\prime}+\mathbf{FW}_{t-1|T}\mathbf{F}^{\prime}]\nonumber \\
+ & \frac{1}{T-1}\sum_{t=2}^{T}\left(\boldsymbol{\theta}_{t|T}-\mathbf{F}\boldsymbol{\theta}_{t-1|T}-\mathbf{U}\mathbf{u}_{t}\right)\left(\boldsymbol{\theta}_{t|T}-\mathbf{F}\boldsymbol{\theta}_{t-1|T}-\mathbf{U}\mathbf{u}_{t}\right)^{\prime}.\label{eq:update_Q}
\end{align}

Finally, the mean of the initial distribution is updated with $\boldsymbol{\mu}^{\ast}=\boldsymbol{\theta}_{1|T}$
from $\frac{\partial}{\partial\boldsymbol{\mu}^{\ast}}q\left(\mathbf{w}^{\ast}|\mathbf{w}\right)=\mathbf{0}$.
The covariance matrix $\boldsymbol{\Sigma}$ for the initial parameters
is fixed in this optimization.

\section{Results}

\subsection{Simulation of a network of 3 neurons}

\begin{figure}[bt] \begin{center} \includegraphics[width=36pc]{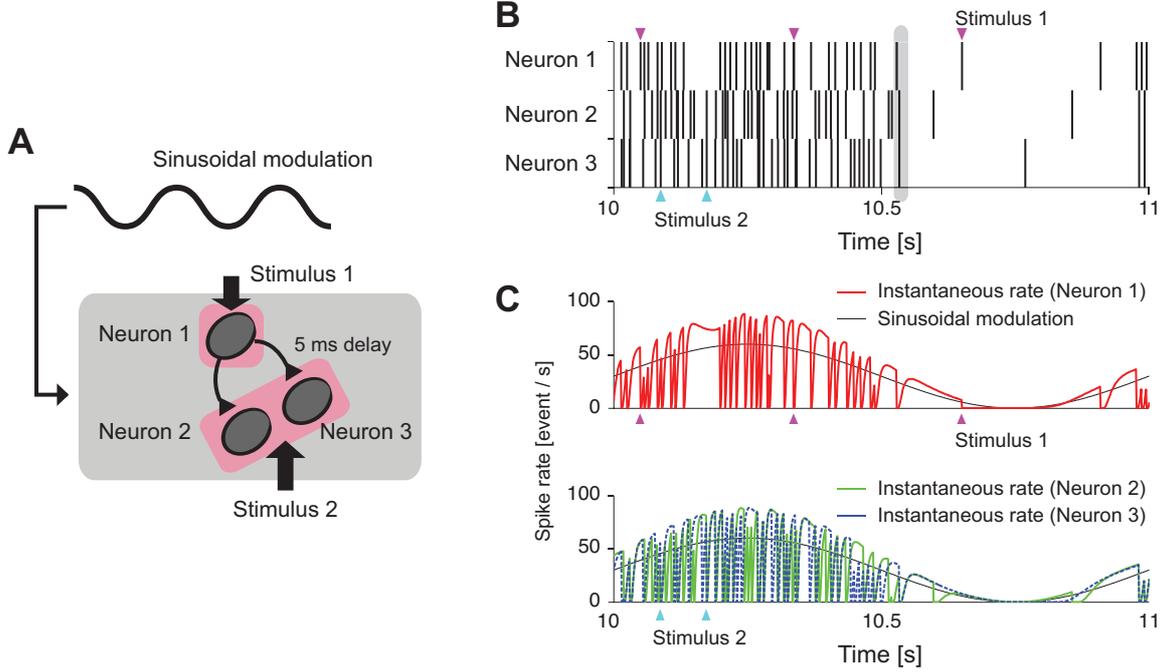}  \end{center} \caption{\label{fig:Figure1}(A) Schematic diagram of a simulated network of 3 neurons. Neuron 1 makes excitatory synaptic contacts to Neuron 2 and 3. Stimulus 1 excites Neuron 1 whereas Stimulus 2 excites Neuron 2 and 3 simultaneously. In addition, all neurons receive sinusoidal rate modulation. (B) Simulated spiking activity of the network. A short period (1 s) of the total 30 s length is shown.  The magenta and cyan triangles represent occurrence times of Stimulus 1 and 2, respectively. The gray bar highlights simultaneous spikes of Neuron 2 and 3 that are causally induced 5 ms after a spike occurs in Neuron 1. (C) Instantaneous spike rates. (Top) The red trace is the instantaneous spike rate of Neuron 1 simulated as an inhomogeneous renewal point process whos instantaneous inter-spike interval is given by the inverse Gaussian distribution ($f\left(t;\kappa\right)=\sqrt{\frac{\kappa}{2\pi t^{3}}}\exp\left[-\frac{\kappa}{2t}\left(t-1\right)^{2}\right]\text{for }x>0\text{, }0\text{ for }x<0$, $\kappa = 1.8$ for all neurons). The inhomogeneous rate is modulated by the sinusoidal function (black solid line, frequency: 1 Hz; mean rate and amplitude: 30 spikes/s). (Bottom) Instantaneous spike rates of Neuron 2 and 3 (solid green line and dashed blue line, respectively). } \end{figure}In
order to test the method, we simulate spiking activity of 3 neurons
that possess specific characteristics in spike generation and connectivity
as follows (See Fig.~\ref{fig:Figure1}A). (1) The instantaneous
firing rate of each neuron model depends on its own spike history
in order to reproduce refractoriness in neuronal spiking activity.
To achieve this, we adopt a renewal point process model whose instantaneous
inter-spike interval (ISI) distribution is given by an inverse Gaussian
distribution as a model of the stochastic spiking activity. (2) The
firing rate of each neuron model varies across time in order to reproduce
the ongoing activity. For that purpose, spike times of each neuron
are generated from the renewal process by adding inhomogeneity to
the underlying rate using the time-rescaling method described in \cite{Brown01}.
The underlying rate of the inhomogeneous renewal point process model
is modulated using a sinusoidal function (frequency: 1 Hz; mean and
amplitude: 30 spikes/s). This rate modulation is common to the 3 neurons.
(3) The neurons are activated by externally triggered stimulus inputs.
To realize the stimulus responses, we deterministically induce spikes
at predetermined timings of the stimuli. We consider two stimuli,
one (Stimulus 1) that induces a spike in Neuron 1, and the other (Stimulus
2) that induces synchronous spikes in Neuron 2 and Neuron 3. The timings
of external stimuli are not related to the sinusoidal time-varying
rate, but randomly selected in the observation period (On average
each stimulus happens once in 1 second). (4) There is feedforward
circuitry in the network. We assume that Neuron 1 makes excitatory
synaptic contacts to Neuron 2 and Neuron 3. To realize this, 5ms after
a spike occurs in Neuron 1, we induce simultaneous spikes in Neuron
2 and Neuron 3 with a probability 0.5. 

We simulate spike sequences with a length of 30 seconds using 1 ms
resolution for numerical time steps (An example of a short period
(1 s) is shown in Fig.~\ref{fig:Figure1}B). Figure \ref{fig:Figure1}C
displays the instantaneous spike-rates (conditional intensity functions
of point processes) of Neuron 1 (Top, red line) and Neuron 2 \& 3
(Bottom, green and blue lines) underlying the spiking activity in
Fig.~\ref{fig:Figure1}B. The black lines indicate sinusoidal rate
modulation common to all neurons. In addition, spikes are induced
in Neuron 1 at the onsets of Stimulus 1 (magenta triangles). Similarly,
simultaneous spikes of Neuron 2 and Neuron 3 are generated at the
onsets of Stimulus 2 (cyan triangles). In the traces of instantaneous
spike-rates in Fig.~\ref{fig:Figure1}C, instantaneous increases
caused by the stimuli and synaptic inputs are not displayed. The instantaneous
spike-rate of a neuron is reset to zero whenever a spike is induced
in that neuron.

\subsection{Selection of a state model}

\begin{figure}[b] \includegraphics[width=18pc]{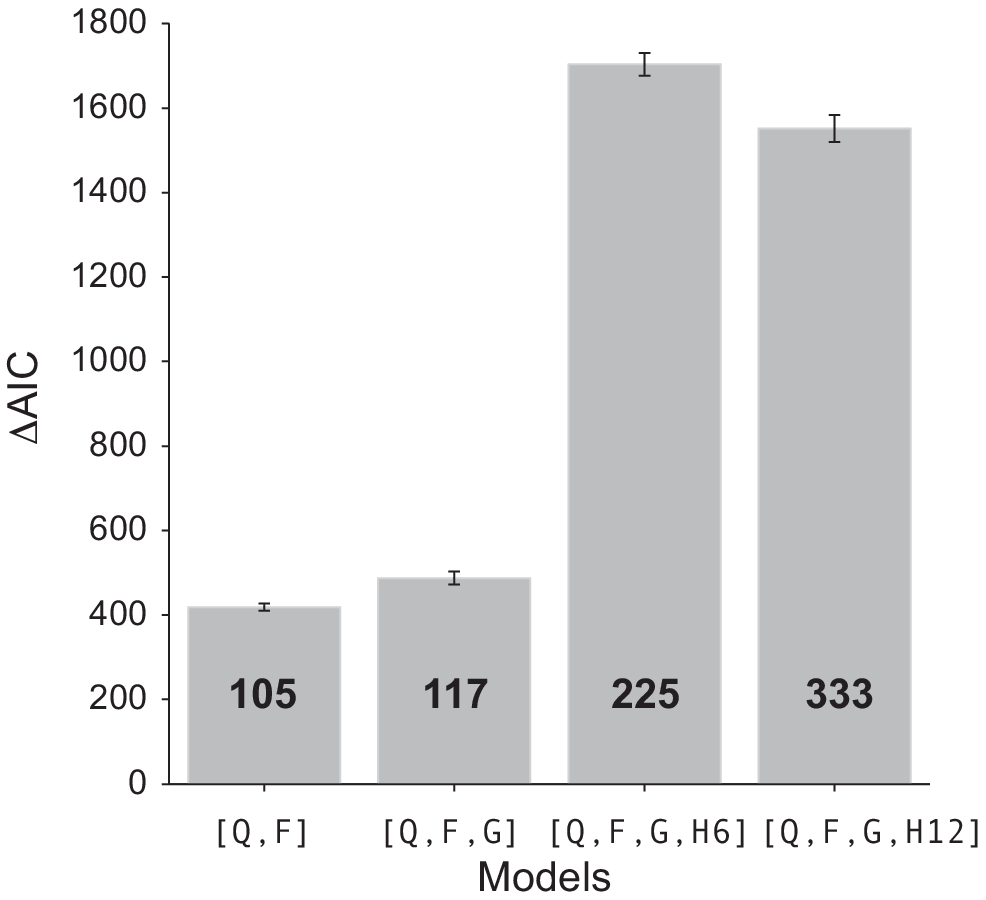}\hspace{2pc} \begin{minipage}[b]{18pc}\caption{\label{fig:Figure2} Comparison of state models by the Akaike Information Criterion (AIC). The state-space models with the following five different state models are comapared: [$\bold Q$], [$\bold Q$, $\bold F$],[$\bold Q$,$\bold F$,$\bold G$], [$\bold Q$,$\bold F$, $\bold G$, $\bold H 6$], and [$\bold Q$,$\bold F$, $\bold G$, $\bold H 12$] (See details of the models for main text). The reduction of the AIC of the last four models from the AIC of the model [$\bold Q$] ($\Delta$AIC) was repeatedly computed for 10 times. The height of the bar indicates the average $\Delta$AIC. The error bar indicates $\pm$ 2 S.E.  The numbers marked on each bar are dimensions of the models (The number of free parameters in the state model). } \end{minipage} \end{figure}We
analyze the simulated ensemble activity by the proposed state-space
model. For this goal, we first construct binary spike patterns, $\mathbf{X}_{1:T}$,
from the simulated spike sequences of 30 seconds (Note: spike times
are recorded in 1 ms resolution) by discretizing them using disjoint
time bins with 2 ms width. We then apply state-space models to the
binary data. The observation model used here contains interactions
up to the second order (a pairwise interaction model):
\begin{align}
p(\mathbf{x}|\boldsymbol{\theta}_{t}) & =\exp\left[\theta_{1}^{t}x_{1}+\theta_{2}^{t}x_{2}+\theta_{3}^{t}x_{3}+\theta_{12}^{t}x_{1}x_{2}+\theta_{13}^{t}x_{1}x_{3}+\theta_{23}^{t}x_{2}x_{3}-\psi(\boldsymbol{\theta}_{t})\right].\label{eq:pairwise_model}
\end{align}
For the state model, we consider 5 different models that include a
set of different components in Eq.~\ref{eq:state_equation}. We select
a model based on the framework of model selection in order to avoid
over-fitting of a model to the data. Details of each state model are
described as follows. 

The first state model assumes $\mathbf{F}=\mathbf{I}$, where $\mathbf{I}$
is an identity matrix, and does not include any of exogenous inputs.
In this model, we optimize only the covariance matrix $\mathbf{Q}$.
The first model is denoted as $[\ensuremath{\mathbf{Q}}]$. The second
state model, denoted as $[\ensuremath{\mathbf{Q}},\ensuremath{\mathbf{F}}]$,
is the first-order auto-regressive model. In this model, we optimize
both the covariance matrix $\mathbf{Q}$ and the auto-regressive parameter
$\mathbf{F}$. The third model, denoted as $[\ensuremath{\mathbf{Q}},\ensuremath{\mathbf{F}},\ensuremath{\mathbf{G}}]$,
additionally includes the stimulus term as exogenous inputs (Stimulus
1 and Stimulus 2). Both the matrix $\mathbf{F}$ and $\mathbf{G}$
are optimized simultaneously in addition to $\mathbf{Q}$. The fourth
model includes both stimulus and spike-history terms. In this model,
the state model includes the history of spiking activity up to the
last 6 time bins ($p=6$). All parameters $\mathbf{F}$, $\mathbf{G}$,
and $\mathbf{H}$ are optimized simultaneously in addition to $\mathbf{Q}$.
This model is denoted as $[\ensuremath{\mathbf{Q}},\ensuremath{\mathbf{F}},\ensuremath{\mathbf{G}},\mathbf{H}6]$.
The structure of the fifth model is the same as the fourth model,
but contains the history of spiking activity up to the last 12 time
bins ($p=12$). The last model is denoted as $[\ensuremath{\mathbf{Q}},\ensuremath{\mathbf{F}},\ensuremath{\mathbf{G}},\mathbf{H}12]$.

In order to select the most predictive model among them, we select
the state-space model that minimizes the Akaike (Bayesian) information
criterion (AIC) \cite{Akaike80}, 
\begin{equation}
\textrm{AIC}=-2l\left(\mathbf{w}^{\ast}\right)+2\textrm{dim}\,\mathbf{w}^{\ast},\label{eq:AIC}
\end{equation}
where $\mathbf{w}^{\ast}$ is the optimized parameter in the Methods
section. The (marginal) likelihood function in Eq.~\ref{eq:AIC}
is obtained by a log-quadratic approximation, i.e, the Laplace method
\cite{Shimazaki12} (See Appendix B for the complete equation). Figure
\ref{fig:Figure2}B displays decreases in AICs ($\Delta\textrm{AIC}$)
of the last four models from the AIC of the first model, $[\ensuremath{\mathbf{Q}}]$.
The larger the $\Delta\textrm{AIC}$ is, the better the state-space
model is expected to predict unseen data. For these data sets, inclusion
of exogenous inputs, in particular the spike history, significantly
decreases the AIC. From this result, we select the state model that
includes the stimulus response term and the spike-history terms up
to the previous 6 time bins.

\subsection{Parameter estimation}

\begin{figure}[tb] \begin{center} \includegraphics[width=36pc]{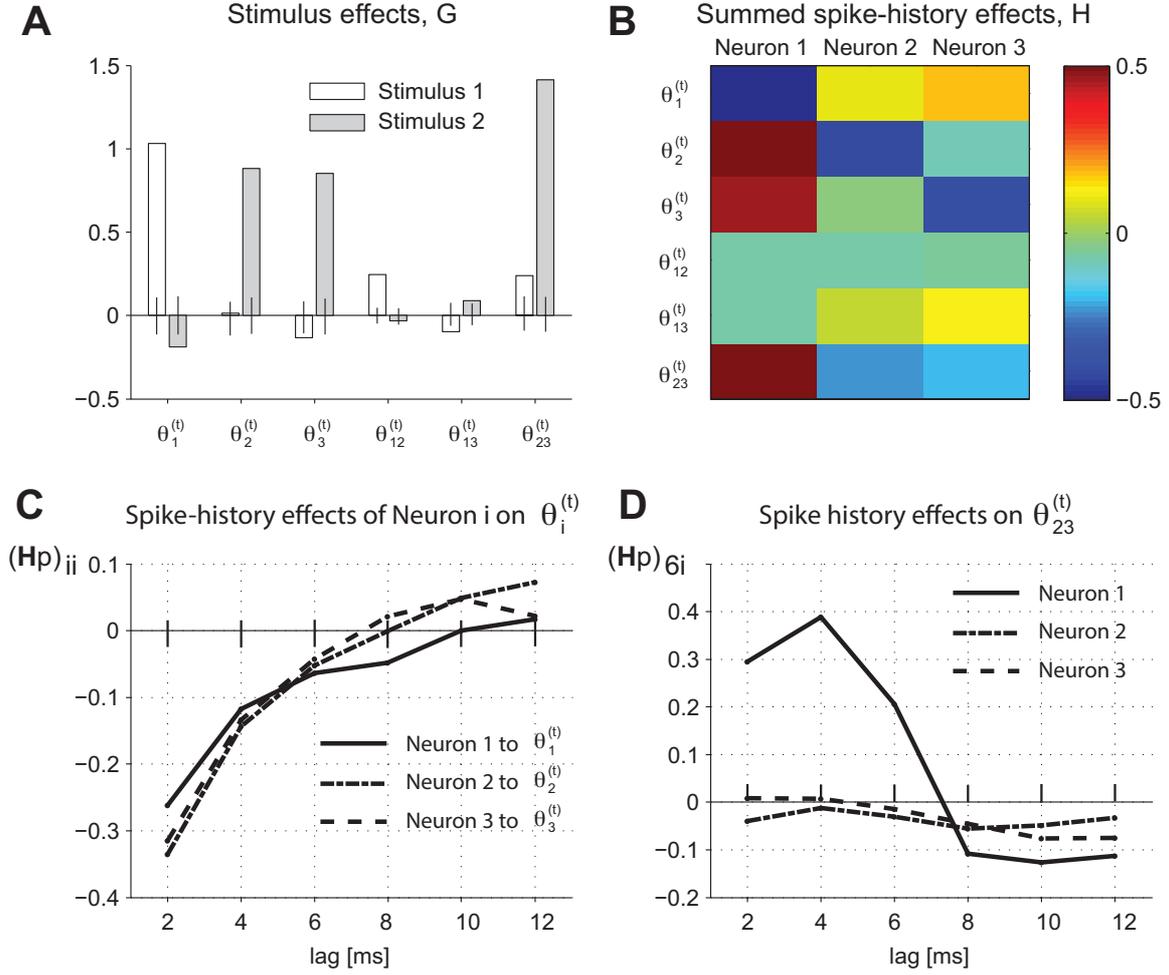} \end{center} \caption{\label{fig:Figure3}Parameter estimation of the state-space model. (A) Effects of Stimulus 1 and 2 on the canonical paramters (the first and second column of $\bold{G}$). The vertical ticks on abscissa indicate the 95\% confidence bounds for each parameters obtained by the surrogate method. (B) Summed spike-history effects. The matrices of spike-history effects, $\bold{H}_{p}$, are summed over the time-lags and shown using color. (C) The effect of a spike occurrence in Neuron $i$ at $p$ time bins before the $t$th bin on $\theta^{(t)}_i$ ($i=1,2,3$). (D) The effect of a spike occurence in Neuron $i$ ($i=1,2,3$) on the interaction parameter $\theta^{(t)}_{23}$.} \end{figure}

We now look at the estimated parameters of the model selected by the
AIC, namely $[\ensuremath{\mathbf{Q}},\ensuremath{\mathbf{F}},\ensuremath{\mathbf{G}},\mathbf{H}6]$.
Due to the limitation in the space, we do not display the estimated
dynamics of the canonical parameters, $\boldsymbol{\theta}_{t}$,
by the recursive Bayesian method (See \cite{Shimazaki09,Shimazaki12}
for the detailed analysis on dynamics of $\boldsymbol{\theta}_{t}$
by this method). The estimated parameters, $\mathbf{G}$ and $\mathbf{H}$,
are summarized in Fig.~\ref{fig:Figure3}. 

Here, in order to test the significance of the estimated parameters,
we construct confidence bounds of the estimates, using a surrogate
method. In this approach, we apply the same state-space model, $[\ensuremath{\mathbf{Q}},\ensuremath{\mathbf{F}},\ensuremath{\mathbf{G}},\mathbf{H}6]$,
to the surrogate data set for the exogenous inputs. In the surrogate
data set, the onset times of external signals (Stimulus 1 and Stimulus
2) are randomized in the observation period. Similarly, we randomly
select $p=6$ bins from the past spiking activity to obtain surrogate
spike history, instead of selecting the last consecutive 6 bins from
time bin $t$. Thus the estimated parameters, $\mathbf{G}$ and $\mathbf{H}$,
are not related to the structure specified in the Section 4.1. We
repeatedly applied the state-space model to the surrogate data (1000
times) to obtain the 95\% confidence bound for the parameter estimation
(vertical ticks in Fig.~\ref{fig:Figure3}A, C and D). 

Figure \ref{fig:Figure3}A displays effects of the two stimuli, $\mathbf{G}$,
on the respective elements in $\boldsymbol{\theta}_{t}$. First, Stimulus
1 significantly increases $\theta_{1}^{(t)}$ whereas changes in the
pairwise interactions by Stimulus 1 are relatively small, indicating
that Stimulus 1 induces spikes in Neuron 1. On the contrary, Stimulus
2 increases to $\theta_{2}^{(t)}$, $\theta_{3}^{(t)}$, and $\theta_{23}^{(t)}$.
In particular, the increase in the interaction parameter $\theta_{23}^{(t)}$
by Stimulus 2 indicates that the presence of Stimulus 2 induces excess
simultaneous spikes in Neuron 2 and Neuron 3 more often than the chance
coincidence expected for the two neurons. 

The spike-history effects are summarized as a summed matrix, $\sum_{i=1}^{p}\mathbf{H}_{i}$,
shown in Fig.~$ $\ref{fig:Figure3}B. Two major effects are observed.
First, the spike history of Neuron $i$ significantly decreases $\theta_{i}^{(t)}$
($i=1,2,3$) (See diagonal of the first $3\times3$ matrix in Fig.~$ $\ref{fig:Figure3}B).
Figure \ref{fig:Figure3}C displays the contribution of a spike in
Neuron $i$ during the previous $p$ time bins to the parameter $\theta_{i}^{(t)}$.
These components primarily, albeit not exclusively, capture the renewal
property of the simulated neuron models. Second, the spike history
of Neuron 1 increases $ $$\theta_{2}^{(t)}$, $\theta_{3}^{(t)}$,
and $\theta_{23}^{(t)}$ (See the first column in Fig.~$ $\ref{fig:Figure3}B),
indicating that spike interactions from Neuron 1 to Neuron 2 \& 3.
In particular, the increase in $\theta_{23}^{(t)}$ due to the spikes
in Neuron 1 during previous 1-3 time bins (Fig.~$ $\ref{fig:Figure3}D)
indicates that the inputs from Neuron 1 induces excess synchronous
spikes in the other two neurons with approximately 2-6 ms delay.

\section{Conclusion}

We developed a parametric method for estimating stimulus responses
and spike-history effects on the simultaneous spiking activity of
multiple neurons when the ensemble themselves exhibit ongoing activity.
The method was tested by simulated multiple neuronal spiking activity
with known underlying architecture. We provided two methods to corroborate
the fitted models. First, based on the result in the preceding paper,
we provided an approximate equation for the log marginal likelihood
(see Appendix B), which was used to select the most predictive state-space
model. Second, we provided a method for obtaining confidence bounds
of the estimated parameters based on a surrogate approach. 

Example spike sequences simulated in this study are overly simplified.
Therefore, the method needs be tested using real neuronal spike data,
e.g., from cultured neurons whose underlying circuit is identified
by electrophysiological studies. In practical applications, it is
recommended to utilize basis functions such as raised cosine bumps
used in \cite{Pillow08} in the exogenous terms in order to capture
the stimulus and spike-history effects with a fewer parameters. In
addition, an appropriate bin size must be selected in order to obtain
a meaningful result in the analysis of real data. Since the bin size
determines a permissible range of synchronous activity, a physiological
interpretation of the result depends on the choice of the bin size.
It is thus recommended to present results based on multiple different
bin sizes in order to confirm a specific hypothesis in a study as
shown in \cite{Riehle00,Shimazaki12}. Methods to overcome an artifact
due to the disjoint binning are discussed in \cite{Gruen99,Pipa08,Gruen09,Hayashi05,Chakraborti11}.
In future, inclusion of such advanced methods will allow us to detect
near-synchronous responses without sacrificing temporal resolution
of the analysis. 

Given that applicability of the method is confirmed in real data,
the proposed method is useful to investigate how ensemble activity
of multiple neurons in a local circuit changes configurations of their
simultaneous responses (synchronous responses) to different stimuli
applied to an animal. Further, it would be interesting to see different
effects of the same stimulus on the ensemble activity when an animal
undergoes different cortical states. 

\ack

The present study is based on the modeling framework developed in
\cite{Shimazaki09,Shimazaki12}. The author acknowledges Prof. Shun-ichi
Amari, Prof. Emery N. Brown, and Prof. Sonja Gr\"un for their support
in construction of the original model. The author also thanks to Dr.
Christopher L. Buckley and Dr. Erin Munro for critical reading of
the manuscript. 

\appendix

\section{Construction of a posterior density by the recursive Bayesian filtering/smoothing
algorithm}

A posterior density of the time-varying $\boldsymbol{\theta}_{t}$,
which specifies the joint probability mass function of spike patterns
at time bin $t$, are obtained by a non-linear recursive Bayesian
estimation method developed in \cite{Shimazaki09,Shimazaki12}. The
method allows us to find a maximum a posteriori (MAP) estimate of
$\boldsymbol{\theta}_{t}$ and its uncertainty, namely the most probable
paths of time-varying canonical parameters $\boldsymbol{\theta}_{t}$
and their confidence bounds given the observed simultaneous activity
of multiple neurons. The estimation procedure completes by a forward
recursion to construct a filter posterior density and then by a backward
recursion to construct a smoother posterior density. In this approach,
the posterior densities are approximated as a multivariate normal
probability density function. 

In the forward filtering step, we first compute mean and covariance
of one step prediction density:
\begin{align}
\boldsymbol{\theta}_{t|t-1} & =\mathbf{F}\boldsymbol{\theta}_{t-1|t-1}+\mathbf{U}\mathbf{u}_{t},\label{eq:one-step_pred_mean}\\
\mathbf{W}_{t|t-1} & =\mathbf{F}W_{t-1|t-1}\mathbf{F}^{\prime}+\mathbf{Q}.\label{eq:one-step_pred_cov}
\end{align}
Then, a mean vector and covariance matrix of the filter posterior
density, which is approximated as a normal density, is given as 
\begin{align}
\boldsymbol{\theta}_{t|t} & =\boldsymbol{\theta}_{t|t-1}+n\mathbf{W}_{t|t-1}(\mathbf{y}_{t}-\boldsymbol{\eta}_{t|t}),\label{eq:filter_mean}\\
\mathbf{W}_{t|t}^{-1} & =\mathbf{W}_{t|t-1}^{-1}+n\mathbf{J}_{t|t},\label{eq:filter_cov}
\end{align}
where $\boldsymbol{\eta}_{t|t}=E\left[\mathbf{f}\left(\mathbf{x}\right)|\mathbf{X}_{1:t},\mathbf{w}\right]$
is the simultaneous spike rates at time bin $t$ expected from the
joint probability mass function, Eq.~\ref{eq:log-linear}, specified
by $\boldsymbol{\theta}_{t|t}$. Thus Eq.~\ref{eq:filter_mean} is
a non-linear equation. We solve Eq.~\ref{eq:filter_mean} by a Newton-Raphson
method. It can be shown that the solution is unique. The matrix $\mathbf{J}_{t|t}$
is a Fisher information matrix of Eq.~\ref{eq:log-linear} evaluated
at $\boldsymbol{\theta}_{t|t}$.

Finally, we compute mean and covariance of a smoother posterior density
as
\begin{align}
\boldsymbol{\theta}_{t|T} & =\boldsymbol{\theta}_{t|t}+\mathbf{A}_{t}\left(\boldsymbol{\theta}_{t+1|T}-\boldsymbol{\theta}_{t+1|t}\right),\label{eq:smooth_mean}\\
\mathbf{W}_{t|T} & =\mathbf{W}_{t|t}+\mathbf{A}_{t}\left(\mathbf{W}_{t+1|T}-\mathbf{W}_{t+1|t}\right)\mathbf{A}_{t}^{\prime}.\label{eq:smooth_covariance}
\end{align}
with $\mathbf{A}_{t}=\mathbf{W}_{t|t}\mathbf{F}^{\prime}\mathbf{W}_{t+1|t}^{-1}$
for $t=T,T-1,\ldots,2,1$. Namely, we start computing Eqs.~\ref{eq:smooth_mean}
and \ref{eq:smooth_covariance} in a backward manner, using $\boldsymbol{\theta}_{T|T}$
and $\mathbf{W}_{T|T}$ obtained in the filtering method at the initial
step. The lag-one covariance smoother, $W_{t-1,t|T}$, is obtained
using the method of De Jong and Mackinnon \cite{DeJong88}:
\begin{align}
\mathbf{W}_{t-1,t|T} & \equiv E[\left.(\boldsymbol{\theta}_{t-1}-\boldsymbol{\theta}_{t-1|T})(\boldsymbol{\theta}_{t}-\boldsymbol{\theta}_{t|T})^{\prime}\right\vert y_{1:T}]=\mathbf{A}_{t-1}\mathbf{W}_{t|T}.\label{eq:lag-one_covariance}
\end{align}

\section{Approximate marginal likelihood function}

The approximated formula of the log marginal likelihood (Eq.~\ref{eq:marginal_loglikelihood})
was obtained in \cite{Shimazaki12} as 
\begin{align}
l(\mathbf{w})\approx & \sum_{t=1}^{T}n\left(\mathbf{y}_{t}^{\prime}\boldsymbol{\theta}_{t|t}-\psi\left(\boldsymbol{\theta}_{t|t}\right)\right)+\frac{1}{2}\sum_{t=1}^{T}\left(\log{\det W_{t|t}}-\log{\det W_{t|t-1}}\right)\nonumber \\
 & -\frac{1}{2}\sum_{t=1}^{T}tr\left[\mathbf{W}_{t|t-1}^{-1}\left(\boldsymbol{\theta}_{t|t}-\boldsymbol{\theta}_{t|t-1}\right)\left(\boldsymbol{\theta}_{t|t}-\boldsymbol{\theta}_{t|t-1}\right)^{\prime}\right].\label{eq:marginal_loglikelihood3}
\end{align}

Here we briefly provide the derivation (See \cite{Shimazaki12} for
details). The log marginal likelihood is written as 
\begin{align}
l(\mathbf{w}) & =\sum_{t=1}^{T}\log p(\mathbf{y}_{t}|\mathbf{y}_{1:t-1},\mathbf{w})=\sum_{t=1}^{T}\log\int p(\mathbf{y}_{t}|\boldsymbol{\theta}_{t})p(\boldsymbol{\theta}_{t}|\mathbf{y}_{1:t-1},\mathbf{w})d\boldsymbol{\theta}_{t}.\label{eq:marginal_loglikelihood2}
\end{align}
The integral in the above equation is approximated as 
\begin{alignat}{1}
\int p(\mathbf{y}_{t}|\boldsymbol{\theta}_{t})p(\boldsymbol{\theta}_{t}|\mathbf{y}_{1:t-1},\mathbf{w})d\boldsymbol{\theta}_{t} & =\frac{1}{\sqrt{(2\pi)^{d}|W_{t|t-1}|}}\int\exp\left[q(\boldsymbol{\theta}_{t})\right]d\boldsymbol{\theta}_{t}\approx\frac{\sqrt{(2\pi)^{d}|W_{t|t}|}}{\sqrt{(2\pi)^{d}|W_{t|t-1}|}}\exp\left[q\left(\boldsymbol{\theta}_{t|t}\right)\right],\label{eq:marginal_loglikelihood_exp}
\end{alignat}
where $q(\boldsymbol{\theta}_{t})=n\left(\mathbf{y}_{t}^{\prime}\boldsymbol{\theta}_{t}-\psi\left(\boldsymbol{\theta}_{t}\right)\right)-\frac{1}{2}\left(\boldsymbol{\theta}_{t}-\boldsymbol{\theta}_{t|t-1}\right)^{\prime}\mathbf{W}_{t|t-1}^{-1}\left(\boldsymbol{\theta}_{t}-\boldsymbol{\theta}_{t|t-1}\right).$
To obtain the second approximate equality, we used the Laplace approximation:
the integral in Eq.~\ref{eq:marginal_loglikelihood_exp} is given
as $\int\exp\left[q\left(\boldsymbol{\theta}_{t}\right)\right]d\boldsymbol{\theta}_{t}\approx\sqrt{(2\pi)^{d}|W_{t|t}|}\exp\left[q\left(\boldsymbol{\theta}_{t|t}\right)\right]$.
Here we note that a solution of $q(\boldsymbol{\theta}_{t})=0$ is
equivalent to the filter MAP estimate, $\boldsymbol{\theta}_{t|t}$.
By applying Eq.~\ref{eq:marginal_loglikelihood_exp} to Eq.~\ref{eq:marginal_loglikelihood2},
we obtain Eq.~\ref{eq:marginal_loglikelihood3}.

\section*{References}
\bibliographystyle{iopart-num}
\bibliography{references}

\end{document}